\documentclass[]{pasj01}
\draft
\begin{document} 
\Received{}
\Accepted{}
\title{Spectral and timing properties of the black hole candidate X1755$-$338 observed in 1989--1995}

\author{
Shigeo \textsc{Yamauchi}\altaffilmark{$\ast$} and 
Sakiho \textsc{Maeda}
}
\altaffiltext{}{Faculty of Science, Nara Women's University, Kitauoyanishimachi, Nara 630-8506}
\email{yamauchi@cc.nara-wu.ac.jp}
\maketitle

\begin{abstract}
We report results of an analysis of the black hole (BH) candidate source X1755$-$338 in 1989, 1990, 
and 1991 with Ginga, and in 1995 with ASCA. 
The spectra were well represented by a model consisting of a soft thermal emission from an accretion disk 
and a hard X-ray tail. 
The normalization of the multi-color disk model, relating to the inner disk radius, 
was similar to each other. 
The unabsorbed X-ray fluxes from the disk component in the 0.01--10 keV band were estimated to be 
1.3$\times$10$^{-9}$, 3.0$\times$10$^{-9}$, 9.8$\times$10$^{-10}$, and 2.4$\times$10$^{-9}$ erg s$^{-1}$ cm$^{-2}$ in 1989, 1990, 1991, and 1995, respectively, 
and are proportional to $kT_{\rm in}^4$, where $kT_{\rm in}$ is 
a temperature at the inner disk radius. 
Based on the standard accretion disk model for a non-rotating BH, 
our results suggest either a small BH mass or a large inclination angle. 
Otherwise, X1755$-$338 is a rotating BH. 
The hard X-ray intensity was found to be variable, while the soft X-ray intensity was stable. 
Although the previous work showed the existence of an iron line at 6.7 keV,
no clear iron line feature was found in all the spectra. 
We infer that most of the iron line flux reported in the previous work was due to 
contamination of the Galactic diffuse X-ray emission. 
\end{abstract}

\section{Introduction}

X1755$-$338 is an X-ray binary source located in the direction of the Galactic bulge, 
exhibiting an ultrasoft spectrum and  a recurrent dip with a period of 4.4h, \citep{Jones1977,WM1984,White1984,Mason1985}.
The spectrum was well represented by a model consisting of a soft thermal emission and a hard X-ray tail \citep{Pan1995,Seon1995}, 
which is well consistent with those commonly found in black hole binaries (BHBs) in the high/soft state 
(e.g., \cite{Tanaka1995}).
Thus, X1755$-$338 is classified into BHB candidates \citep{WM1984,White1984,Tanaka1995}.
The optical counterpart was identified with a star with $V$=18 mag and 
the distance, $D$, is thought to be less than 9 kpc \citep{Mason1985},  
while \citet{Wachter1998} indicated that $D$ is larger than 4 kpc. 
Assuming that X1755$-$338 is a non-rotating black hole (BH, Schwartzshild BH), 
\citet{Pan1995} and \citet{Seon1995} pointed out that the derived inner disk radius is small compared 
with typical BHBs.  
\citet{Pan1995} argued that if X1755$-$338 is a BHB accompanied 
with an accretion disk with the inner disk radius of 9 km, 
the inclination angle should be $>$70$^{\circ}$. 
However, it is inconsistent with the inclination angle of  $<$70$^{\circ}$ suggested to avoid eclipses by the companion \citep{White1984}. 

\citet{Seon1995} reported that an emission line at $\sim$6.7 keV was detected in 1991. 
The authors argued that the emission region was close to the central source 
because the line energy (6.7 keV) is higher than that of the fluorescence line from the cold matter (6.4 keV). 
It is known that Galactic diffuse X-ray emission (GDXE), unresolved X-ray emission in the Galaxy, 
has an intense 6.7 keV line from highly ionized iron 
and is located not only along the Galactic plane and in the Galactic center 
but also in the Galactic bulge region \citep{Yamauchi1993,Yamauchi2016}.
Thus, whether the iron line exists or not should be carefully examined.

Although X1755$-$338 has been a persistent bright X-ray source since its discovery, 
it suddenly turned off in 1996 \citep{Roberts1996}. 
During the low state, a filamentary structure around X1755$-$338 was discovered \citep{Angelini2003}. 
This is thought to be a jet from X1755$-$338, indicating a BHB. 
Recently, it was reported that X1755$-$338 entered an outburst phase \citep{Moreminsky2020,Draghis2020}. 

The nature of X1755$-$338 has not been well understood. 
In this paper, we present results of the spectral and timing analyses using data obtained with Ginga and ASCA 
in 1989--1995, before entering the low state, 
and discuss the properties.

\section{Data analysis and results}

\subsection{Ginga data}

Ginga observations were carried out with the Large Area Counters (LAC).
The total effective area and the energy range of the LAC were 4000 cm$^2$ and 1--37 keV, respectively. 
The FWHM of the LAC field of view (FOV) was about 1$^{\circ}$ and 2$^{\circ}$ along and perpendicular to the scan path, respectively.
The accuracy of the satellite attitude determination was better than \timeform{0.D1}.
Details concerning Ginga and the LAC were given in \citet{Makino1987} and \citet{Turner1989}, respectively.

We excluded data obtained at the South Atlantic Anomaly, 
during earth occultation, at the low elevation angle from the earth rim 
of $<$ 5$^{\circ}$, and in the high background regions at 
low geomagnetic cut-off rigidity of $<$9 GV.
The non-X-ray background (NXB) was subtracted using a method given by \citet{Awaki1991}.

\subsubsection{Scanning observations in 1989}

The scanning observations in the Galactic bulge region were carried out on 1989 April 9.
The scan path was along the Galactic plane ($b\sim-$\timeform{6.D2}) 
and the scan speed was 0.95 deg min$^{-1}$.
Data were taken in MPC-2 mode.
Scan data with a high background counts were excluded. 
Although multiple scans were carried out around X1755$-$338 in the $l$=355$^{\circ}$--0$^{\circ}$ region, 
two scan data, in which X1755$-$338 was clearly detected, were picked up 
and were added in order to get a high signal-to-noise ratio.

Figure 1 shows the scan profiles in the 1.2--2.9, 2.9--6.4, 6.4--18.5, and 1.2--18.5 keV energy bands. 
A soft X-ray source was clearly found at the scan angle of $\sim -$3$^{\circ}$. 
In order to determine the source position, 
we applied the same procedure as that described in \citet{Yamauchi2005}.
We fitted the scan profile with a model consisting of contribution of point sources and the sky background, 
the Cosmic X-ray background (CXB) and the GDXE.
The sky background was assumed to be constant along the scan path.
We applied the scan fitting to the 1.2--18.5 keV energy band data. 
Since positive residuals were seen at the scan angle of $\sim -$4$^{\circ}$, we added another point source. 
As a result, the scan angles of the sources were determined to be \timeform{-2.D76}$\pm$\timeform{0.D10} 
and \timeform{-3.D77}$\pm$\timeform{0.D22}:  
the former was identified with X1755$-$338.  

Taking into account the calibration uncertainty of 
the collimator response function ($\le$3\%, \cite{Kondo1988}),
intensity variation of point sources during the scanning observation,
contribution of unresolved faint sources on the Galactic plane, and local structure of the GDXE, 
the model represented the global profile ($\chi^2$/d.o.f.=22.5/16) and the source positions were determined well. 

We, next, made a scan profile in each energy channel and fitted them with the same model function, 
fixing the scan angles of the sources and freeing the intensity.
Using the peak intensities of X1755$-$338 derived from the scan fitting,
we made the X-ray spectrum. 
Figure 2(a) shows the X-ray spectrum obtained in 1989 using this method.
Taking into account the elevation angle of X1755$-$338 from the scan path, 
we made the aspect correction. 

\begin{figure}
  \begin{center}
    \includegraphics[width=8cm]{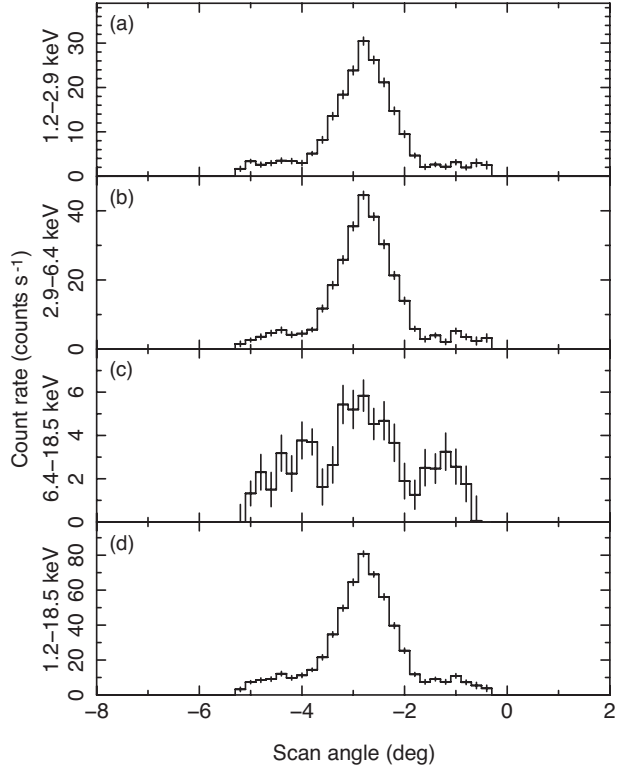}
  \end{center}
  \caption{Scan profile obtained with the Ginga LAC in 1989 April. (a) 1.2--2.9 keV, (b) 2.9--6.4 keV, and (c) 6.4--18.5, and (d)1.2--18.5 keV energy bands. 
  The scan angle of 0$^{\circ}$ is referred to $l$=\timeform{0.D0}. Errors of the data points are at the 1$\sigma$ level. 
  }\label{fig:img}
\end{figure}

\subsubsection{Pointed observations in 1990 and 1991}

\begin{figure*}
  \begin{center}
      \includegraphics[width=8cm]{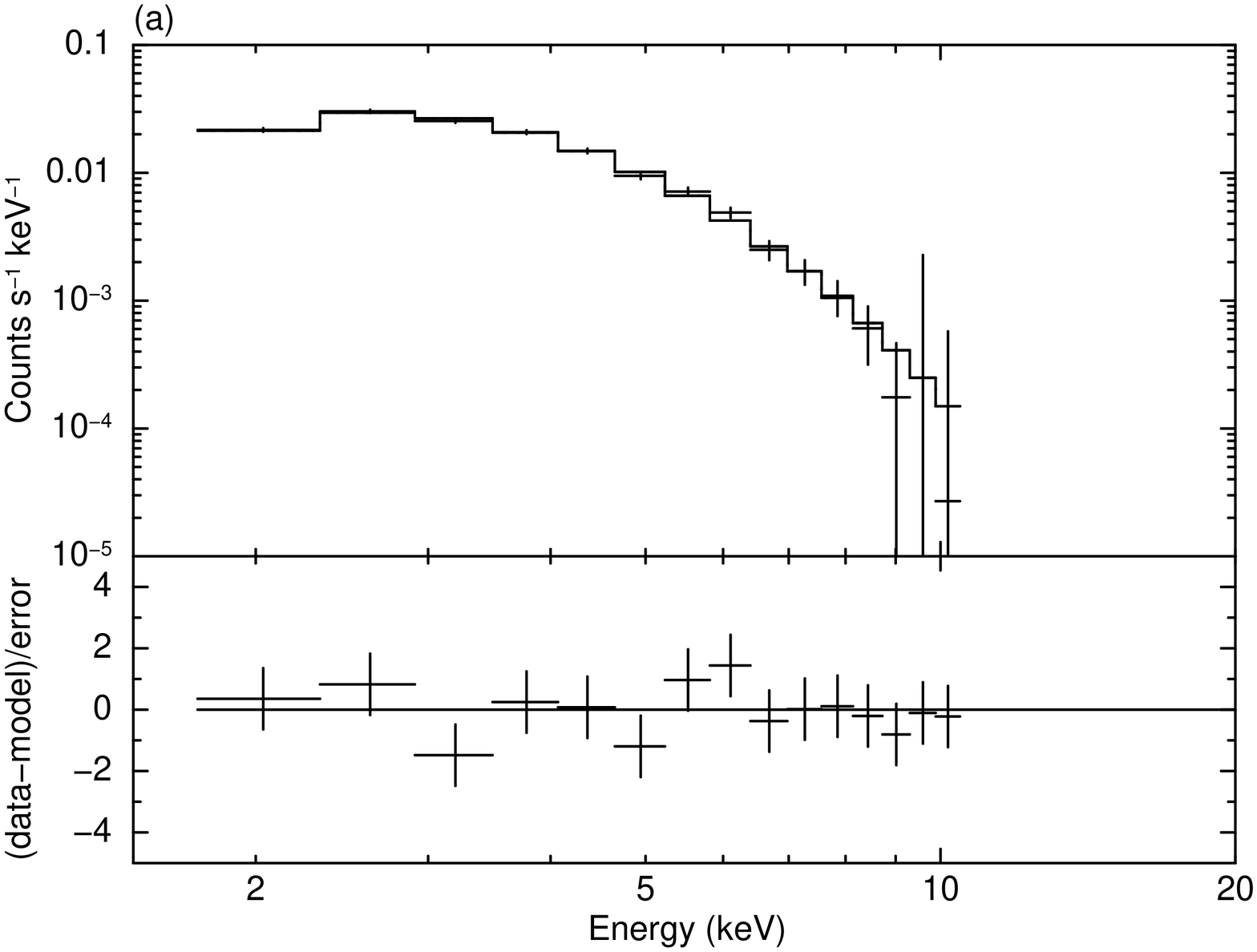}
      \includegraphics[width=8cm]{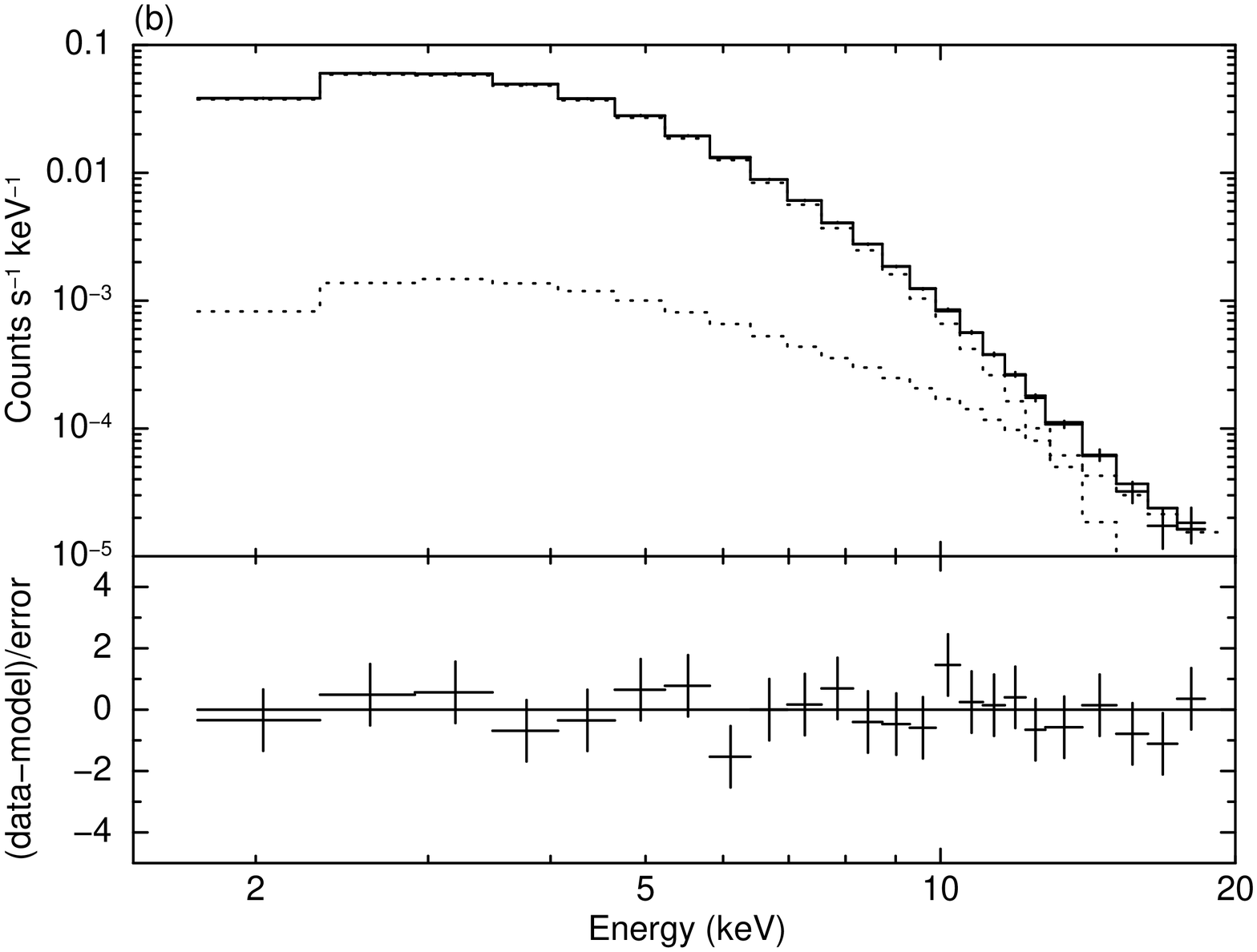}
      \includegraphics[width=8cm]{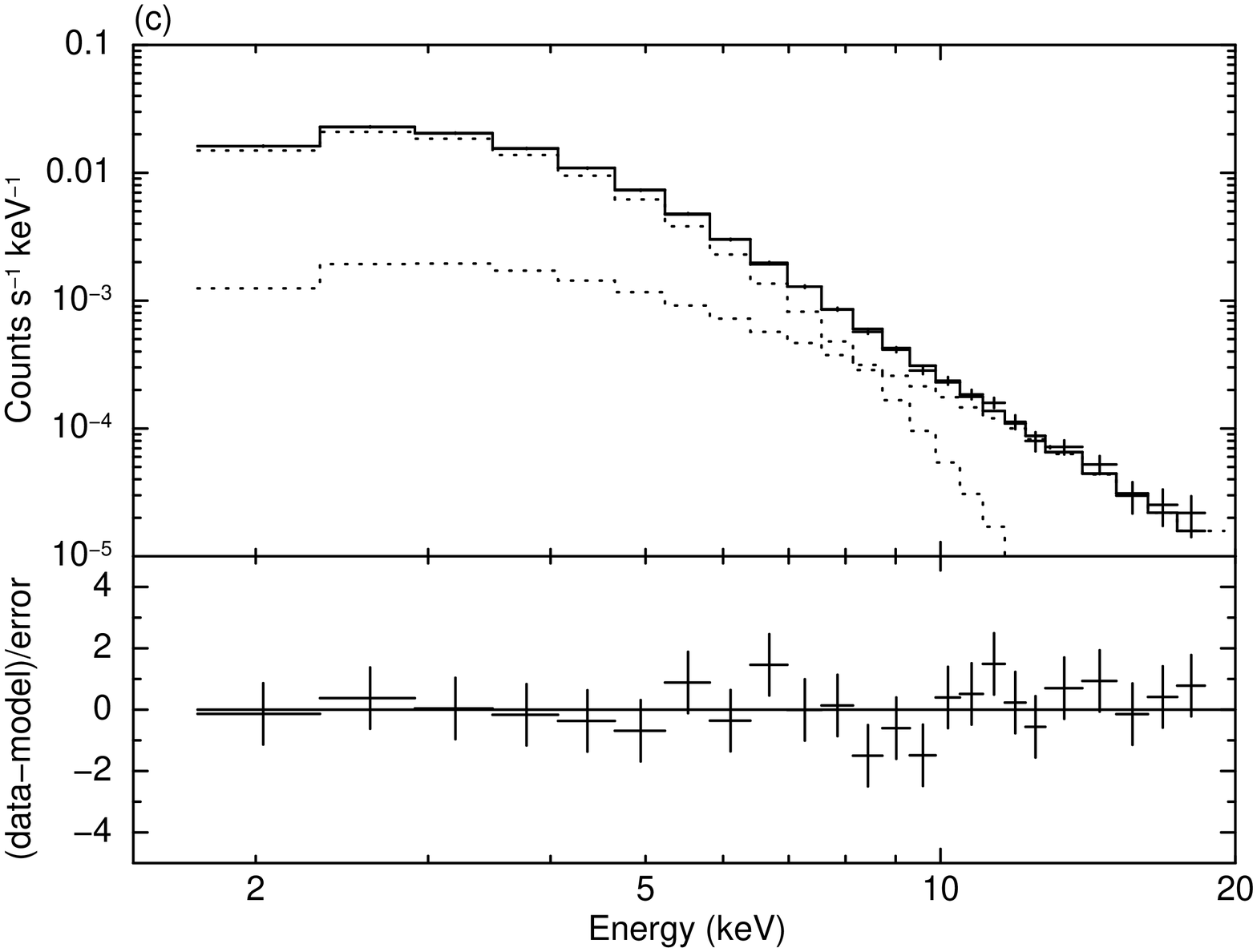}
      \includegraphics[width=8cm]{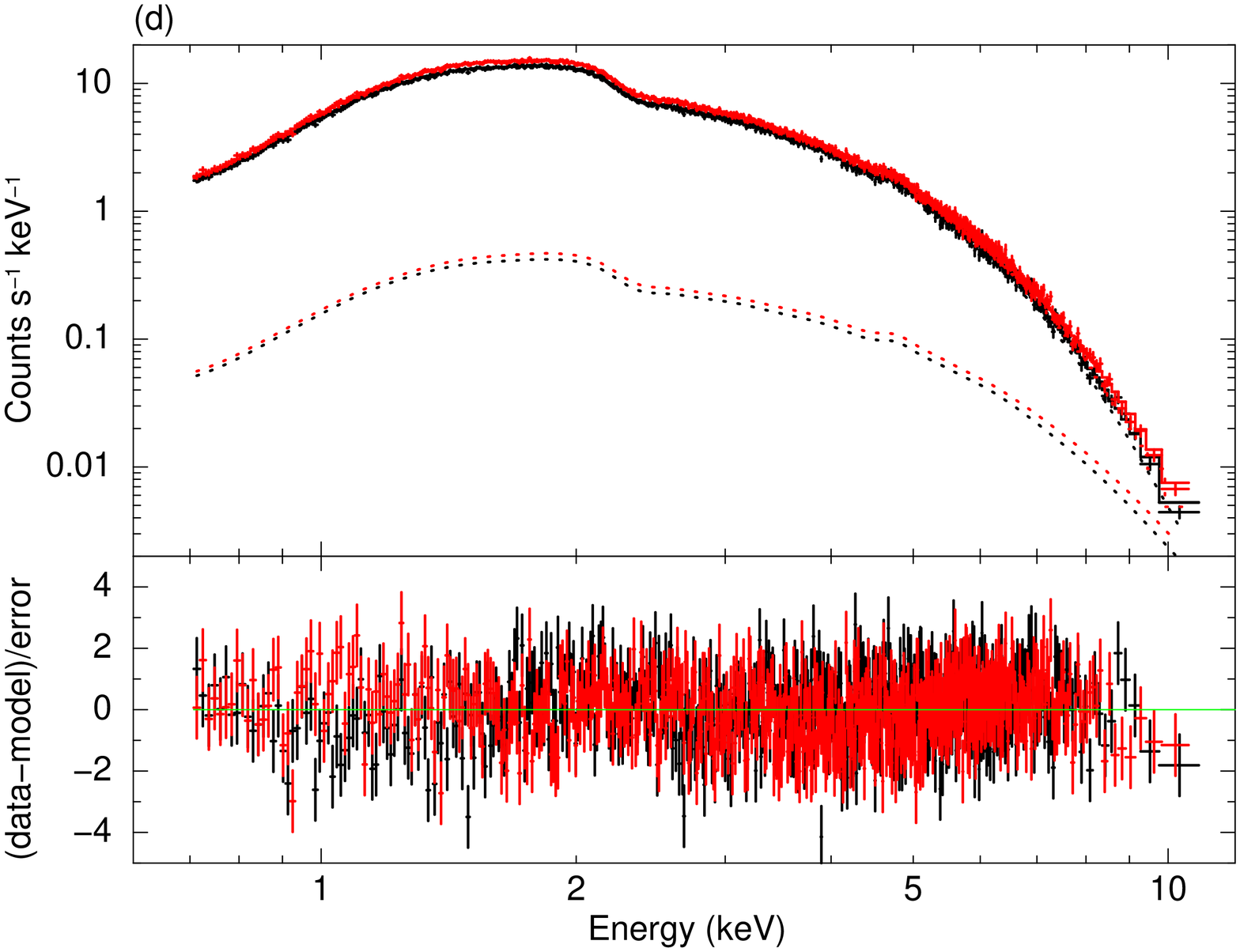}
  \end{center}
  \caption{Spectra of X1755$-$338, 
  (a) 1989 April 9 (Ginga), (b) 1990 April 24--25 (Ginga), (c) 1991 September 22-23 (Ginga), and (d) 1995 March 14--15 (ASCA, black: GIS2 and red: GIS3). 
  Crosses and histograms in the upper panel for each figure are data points and the best-fit model, respectively. 
  The lower panel for each figure shows the residuals from the best-fit multi-collor disk and thermal Comptonized continuum model. 
  Errors of the data points are at the 1$\sigma$ level. 
}\label{fig:img}
\end{figure*}

%
\begin{table*}[t]
\caption{The best-fit parameters of spectral analysis for the non-dip period.}
\begin{center}
\begin{tabular}{llcccc} \hline  
Component & Parameter &\multicolumn{4}{c}{Value$^{\ast}$}\\ 
& & Ginga 1989 & Ginga 1990 & Ginga 1991 & ASCA 1995\\ \hline 
\multicolumn{6}{c}{Model 1: (multi-color disk$+$power-law)$\times$absorption} \\ \hline
phabs& $N_{\rm H}$ ($\times10^{21}$ cm$^{-2}$) &	 
$<$3.1 & 5.6$\pm$0.6 & 3.8$\pm$1.0 & 3.6$^{+0.1}_{-0.2}$\\
diskbb & $kT_{\rm in}^{\dag}$ (keV) &  			
1.09$^{+0.04}_{-0.07}$  & 1.24$^{+0.02}_{-0.01}$ & 0.99$\pm$0.02 & 1.19$\pm$0.01\\
 &  ($r_{\rm in}$/$d_{\rm 10}$)$^2$ cos\ $\theta$$^{\ddag}$& 
 42$^{+14}_{-7}$ & 57$\pm$4 & 46$\pm$5 & 55$\pm$2  \\ 
Power law& $\Gamma$ & 					
2.5 (fixed) & 2.5 (fixed) & 2.5 (fixed) & 2.5 (fixed) \\
& Normalization$^{\S}$ & 					
$<$0.09 & 0.065$\pm$0.07 & 0.068$^{+0.006}_{-0.005}$ & 0.064$^{+0.012}_{-0.014}$ \\ 
& $\chi^2$/d.o.f. & 							
8.3/11 & 14/22  & 16/22  & 1147/1068    \\
\hline 
\multicolumn{6}{c}{Model 2: (multi-color disk$+$thermal comptonized continuum)$\times$absorption} \\ \hline
phabs& $N_{\rm H}$ ($\times10^{21}$ cm$^{-2}$) &	
 $<$2.4 & 5.0$\pm$0.6 & 2.6$\pm$1.1 & 3.2$\pm$0.1\\
diskbb & $kT_{\rm in}^{\dag}$ (keV) &  			
1.10$^{+0.03}_{-0.09}$  & 1.24$\pm$0.01 & 0.98$\pm$0.02 & 1.18$\pm$0.01\\
 &  ($r_{\rm in}$/$d_{\rm 10}$)$^2$ cos\ $\theta$$^{\ddag}$& 					
 42$^{+16}_{-7}$ & 58$\pm$4 & 48$\pm$6 & 58$\pm$2  \\ 
 & Flux$^{\sharp}$ ($\times$10$^{-9}$ erg s$^{-1}$ cm$^{-2}$)& 1.3& 3.0& 0.98& 2.4\\
nthcomp & $kT_{\rm 0}$ (keV) & link to $kT_{\rm in}$ & link to $kT_{\rm in}$ & link to $kT_{\rm in}$ & link to $kT_{\rm in}$ \\
 &  $\Gamma$ & 						2.5 (fixed) & 2.5 (fixed)  & 2.5 (fixed) & 2.5 (fixed) \\
  & $kT_{\rm e}$ (keV) & 100 (fixed) & 	100 (fixed) & 100 (fixed) & 100 (fixed) \\
& Normalization & 					
$<$0.02 & 0.009$\pm$0.002 & 0.013$\pm$0.002 & 0.012$\pm$0.003 \\ 
 & Flux$^{\sharp}$ ($\times$10$^{-10}$ erg s$^{-1}$ cm$^{-2}$)& $<$1.5& 0.87& 1.1& 1.1\\
& $\chi^2$/d.o.f. & 		8.4/11 & 14/22  & 17/22  & 1151/1068    \\
\hline \\\end{tabular}
\end{center}
\vspace{-12pt}
$^{\ast}$ The quoted errors are in the 90\% confidence limits. \\
$^{\dag}$ Temperature at an inner-disk radius of the accretion disk ($r_{\rm in}$).\\
$^{\ddag}$ $d_{\rm 10}$ is a distance to the source in unit of 10 kpc and $\theta$ is an inclination angle.\\
$^{\S}$ The unit is photons s$^{-1}$ cm$^{-2}$ keV$^{-1}$ arcmin$^{-2}$ at 1 keV.\\
$^{\sharp}$ Unabsorbed flux in the 0.01--10 keV band. 
\end{table*}

The pointed observations of X1755$-$338 were made on 1990 April 24--25 and 1991 September 22--23. 
Data were taken in MPC-2 and MPC-1 mode in the 1990 and 1991 observations, respectively. 
The time resolution was 0.5 s (medium bit rate) or 2.0 s (low bit rate) in 1990 and 16 s (low bit rate) in 1991. 
The data were the same as those used in \citet{Seon1995}.

As reported by \citet{Seon1995}, the dip feature was seen in the 1990 and 1991 observations. 
Here, we made spectra in only the non-dip period. 
In order to obtain a high signal-to-noise ratio, 
we made a spectrum from the data with a collimator transmission of $>$90\% in the 1990 observation. 
The satellite attitude during the pointed observation in 1991 was unstable, 
and hence we picked up the data with a collimator transmission of $>$80\%. 
The averaged pointed positions were ($l$, $b$)=(\timeform{357.D3}, \timeform{-4.D9}) 
and ($l$, $b$)=(\timeform{357.D4}, \timeform{-4.D7}) in the 1990 and 1991 observations, respectively, while 
the exposure times are 9.3 ks and 6.4 ks, respectively.

The sky background data were taken at the pointed position of 
($l$, $b$)=(\timeform{358.D7}, \timeform{-4.D3}), close to X1755$-$338, in 1990 April 24
and the NXB was subtracted from the data. 
Figure 2(b) and 2(c) show X-ray spectra after the NXB and sky background subtraction obtained in the 1990 and 1991 observations, 
respectively. 
Using the averaged satellite attitude, we made the aspect correction. 

In order to examine timing property, we made light curves.  
After the subtraction of the NXB and the sky background, the aspect correction was made. 

\subsection{ASCA data}

The field centered on ($l$, $b$)=(\timeform{357.D282}, \timeform{-4.D896}) 
was observed on 1995 March 14--15 with the two Solid-state Imaging 
Spectrometers (SIS0, SIS1) and the 
two Gas Imaging Spectrometers (GIS2, GIS3) placed at the focal plane 
of the thin foil X-ray Telescope (XRT) onboard ASCA. 
Details of ASCA and the instruments are given in separate papers 
(ASCA satellite: \cite{Tanaka1994}; XRT: \cite{Serlemitsos1995}; 
SIS: \cite{Burke1991}; GIS: \cite{Makishima1996, Ohashi1996}).
Since X1755$-$338 was bright, pile up phenomena and telemetry overflow were occurred.
To avoid these data, we used only the GIS data with a high bit rate.

The GIS was operated in PH mode and 64 pixel mode. 
The FOV was $\sim$50$'$ in diameter. 
We excluded the data obtained at the South Atlantic Anomaly, during the earth occultation, 
at the low elevation angle from the earth rim of $<$ 5$^{\circ}$, 
and in the high background regions at low geomagnetic cut off rigidities of $<$6 GV.  
We also applied a rise-time discrimination technique to reject particle events.
After the selection, the exposure time is 19.6 ks. 
The source spectrum was extracted from a circle with a radius of $\sim$4$'$, 
while the background spectrum was accumulated from the outer region in the same FOV.
Figure 2(d) shows the X-ray spectrum after the background subtraction.

In order to examine timing property, we made light curves.  
After the subtraction of the NXB and the sky background, the data obtained with the GIS 2 and GIS 3 were added.

\subsection{Spectrum}
 
As shown in figure 2, X1755$-$338 exhibits thermal dominant spectra in all the periods, 
which is commonly seen in BHBs in high/soft state (e.g., \cite{Tanaka1995}).
Thus, we applied a multi-color disk model ({\tt diskbb} in XSPEC, \cite{Mitsuda1984}) with low-energy absorption by interstellar matter (ISM) ({\tt phabs} in XSPEC). 
The cross sections of the photoelectric absorption were taken from \citet{bcmc1992}. 
The model well represented the low-energy band spectra, but the positive residuals were remained in the high energy band, and hence
we added a power law function as a hard tail component. 
The model formula is 
\begin{equation}
{\rm Model\ 1}: ({\rm diskbb+power\ law})\times{\rm Abs}_{\rm ISM}.
\end{equation}

The best-fit photon index was in the range of 2.4--2.6, but was not well constrained, and hence the photon index was fixed to be 2.5.  
The best-fit parameters are listed in table 1.

The power law model is not a physical but an empirical model. 
The hard tail component is considered as the Compton-scattered X-rays by high energy electrons 
in a hot plasma near the central source (e.g., \cite{Tanaka1995}). 
Thus, instead of the power law model, we applied a thermal Comptonized continuum model ({\tt nthcomp} in XSPEC, \cite{Zdziarski1996,Zycki1999}). 
The model formula is 
\begin{equation}
{\rm Model\ 2}: ({\rm diskbb+nthcomp})\times{\rm Abs}_{\rm ISM}.
\end{equation}

The temperature of seed photons was linked to the temperature at the inner-disk radius, $kT_{\rm in}$,  
while an electron temperature and an asymptotic power law photon index were assumed to be 100 keV and 2.5, respectively. 
This model also gave acceptable fits. 
The best-fit parameters and the observed X-ray flux for each component are listed in table 1, while the best-fit models are plotted in figure 2. 
Figure 3 is a correlation plot between $kT_{\rm in}$ and the unabsorbed energy flux of the disk component, 
which clearly shows that the flux of the disk component is proportional to $kT_{\rm in}^4$

We also examined the electron temperature of 40 keV, 
but the parameters of the multi-color disk model were the same within the statistical errors.

\begin{figure}
  \begin{center}
    \includegraphics[width=8cm]{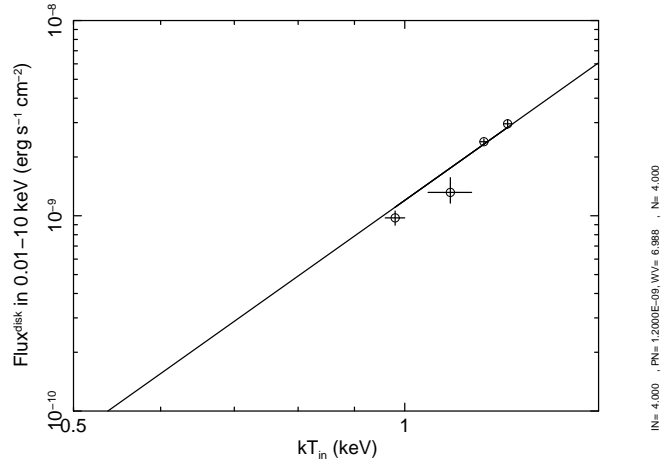}
  \end{center}
  \caption{Correlation plot between a temperature at the inner disk radius and an unabsorbed energy flux of the disk component in the 0.01--10 keV band. 
  The solid line shows the {\it Flux} $\propto kT_{\rm in}^4$ relation. Errors of the data points are at the 1$\sigma$ level. 
}\label{fig:img}
\end{figure}

\subsection{Iron line feature}

\citet{Seon1995} reported that an emission line at $\sim$6.7 keV was detected in the 1991 observation:   
the iron line flux derived from the multi-color disk$+$Comptonized blackbody$+$Gaussian model was 
(2.05$^{+0.74}_{-0.71}$)$\times$10$^{-4}$ photons s$^{-1}$ cm$^{-2}$.
However, in our spectra, no clear line feature is seen (see figure 2). 
We re-fitted the spectrum with model 2 adding a Gaussian line at 6.7 keV 
and obtained upper limits of the line flux (90\% confidence level)
of 1.0$\times$10$^{-3}$, 1.6$\times$10$^{-4}$, 1.7$\times$10$^{-4}$, 
and 2.5$\times$10$^{-4}$ photons s$^{-1}$ cm$^{-2}$ in 1989, 1990, 1991, and 1995, respectively. 

We infer that the iron line flux previously reported in \citet{Seon1995} was mainly due to contamination of the GDXE. 
Although \citet{Seon1995} stated that the sky background data obtained 7 days after the observation in 1991 were used 
(the position of the sky background was not shown),  
no data with the position close to X1755$-$338 were found in 1991 September--October in the Ginga observation logs. 
The GDXE in the Galactic bulge region also has an intense 6.7 keV line from highly ionized iron 
\citep{Yamauchi1993,Yamauchi2016}. 
If the position of the sky background data was not close to X1755$-$338, the GDXE contamination is not negligible. 

The iron line flux in the nearby sky, from which the sky background for this analysis was extracted, 
was determined to be (1.4$\pm$0.5)$\times$10$^{-4}$ photons s$^{-1}$ cm$^{-2}$, 
which is well consistent with those in \citet{Yamauchi1993}. 
Furthermore, the Suzaku results show that the flux of the iron emission line around 
($l$, $b$)$\sim$(\timeform{0D}, \timeform{-5D}) is 
(1--2)$\times$10$^{-8}$ photons s$^{-1}$ cm$^{-2}$ arcmin$^{-2}$ \citep{Yamauchi2016}, 
which corresponds to $\sim$(0.7--1.4)$\times$10$^{-4}$ photons s$^{-1}$ cm$^{-2}$ in the Ginga LAC FOV. 
Thus, the iron line flux of the GDXE is comparable to that reported in \citet{Seon1995}. 

\subsection{Timing property}

\begin{figure}
  \begin{center}
    \includegraphics[width=8cm]{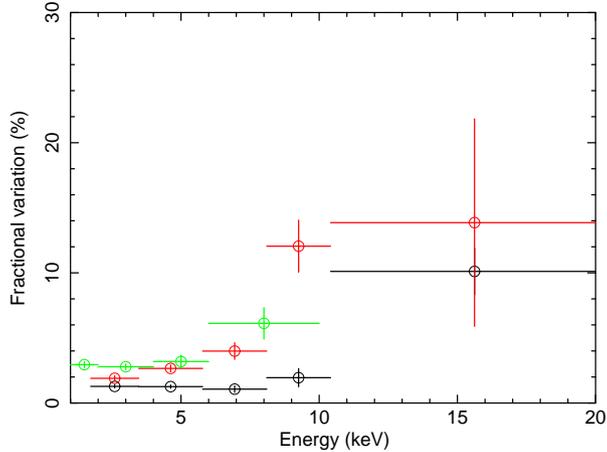}
  \end{center}
  \caption{Fractional variation as a function of the photon energy. 
  The black, red, and green colors show the Ginga data in 1990, 1991, and ASCA data in 1995, respectively. 
  The time-bin-width is 64 s. Errors of the data points are at the 1$\sigma$ level. 
}\label{fig:img}
\end{figure}

Based on the spectral results, 
most of X-rays in the soft X-ray band originate from the accretion disk, 
while those above 8--10 keV come from hard X-ray tail component. 
In order to examine timing property, 
we estimated the fractional variation using the pointed observation data (Ginga data in 1990 and 1991 and the ASCA data in 1995).
The fractional variation is defined as 

\begin{equation}
F_{\rm var}=\frac{\sqrt{S^2-\bar{\sigma}_{\rm err}^2}}{\bar{X}}, 
\end{equation}
where $S^2$, $\bar{\sigma}_{\rm err}^2$, and $\bar{X}$ are the variance of the light curve, the mean error squared, and the mean count rate, 
respectively (e.g., \cite{Edelson2002}). 
X-ray counts were accumulated in the time-bin-width of 64 s.  
The results are shown in figure 4 (the RMS spectrum). 
The variability depends on the energy band: the variability is high in the hard X-ray band.

\section{Discussion and conclusion}

X-ray spectra observed with Ginga and ASCA in 1989--1995 were well represented 
by a soft thermal emission from the accretion disk and a hard tail component. 
A temperature of the inner disk radius $kT_{\rm in}$ is $\sim$1 keV, which are very similar to those of BHBs 
(e.g., \cite{Tanaka1995}). 
Furthermore, $kT_{\rm in}$ is consistent with those of the previous works \citep{Pan1995,Seon1995}. 
Figure 3 shows that the flux of the disk component is proportional to $kT_{\rm in}^4$, 
which is in agreement with those of the standard disk model of BHBs in the high/soft state 
(e.g., \cite{Makishima2000,Kubota2001}). 
The results in \citet{Pan1995} show also the same tendency. 
In addition, the intensity variation shows that the harder X-ray photons are variable (figure 4), 
which suggests that X-rays from an accretion disk are stable and those from a hard X-ray tail are variable (see also figure 2), 
This is also similar to those found in BHBs (e.g., \cite{Ebisawa1991}). 
Thus, X1755$-$338 is most likely to be a BHB. 

The normalization of the {\tt diskbb} model corresponds to ($r_{\rm in}$/$d_{\rm 10}$)$^2$ cos\ $\theta$, 
where $r_{\rm in}$ is an apparent  inner-disk radius of the accretion disk, 
$d_{\rm 10}$ is a distance to the source in unit of 10 kpc, and $\theta$ is an inclination angle. 
We found that the normalization is similar to each other, and hence the inner-disk radius is constant.
Referring to \citet{Kubota1998}, we estimate the BH mass 
with the assumption of a non-rotating BH (the Schwartzshild BH). 
The apparent inner-disk radius, $r_{\rm in}$, is converted to the realistic radius, $R_{\rm in}$,  
by the following formula, 
\begin{equation}
R_{\rm in}=\xi \cdot \kappa^2 \cdot r_{\rm in},
\end{equation}
where $\xi=\sqrt{\displaystyle{\frac{3}{7}}} \cdot \displaystyle{(\frac{6}{7}})$$^3$ \citep{Kubota1998} 
and $\kappa$ (a spectral hardening factor) of 1.7--2.0 \citep{Shimura1995}.
Taking into account the errors, the normalization of the {\tt diskbb} model, 
($r_{\rm in}$/$d_{\rm 10}$)$^2$ cos\ $\theta$, is 
in the range of 36--61 (table 1). 
Using $\kappa=$1.7--2.0, the distance of 4--9 kpc, and the inclination angle of $<$70$^{\circ}$ \citep{White1984}, 
and assuming the last stable orbit of $3 R_{\rm s}=R_{\rm in}$ (e.g., \cite{McClintock2006}), where $R_{\rm s}$ is the Schwartzshild radius, 
we calculated the BH mass, $M_{\rm BH}$, to be $M_{\rm BH}<2.2 M_{\odot}$. 
Adopting $M_{\rm BH}=3 M_{\odot}$, the inclination angle should be $>$79$^{\circ}$. 

The luminosity of the disk component, $L_{\rm disk}$, is estimated to be 
$L_{\rm disk}=$(1.3--4.2)$\times$10$^{37} $($D$/9\ kpc)$^2(\cos\ i/\cos 70^{\circ})^{-1}$. 
This corresponds to $\sim$0.03--0.1 $(\cos\ i/\cos 70^{\circ})^{-1} L_{\rm Edd}$ at a distance of 9 kpc 
or $\sim$0.006--0.02 $(\cos\ i/\cos 70^{\circ})^{-1} L_{\rm Edd}$ at a distance of 4 kpc by assuming a $3M_{\odot}$ BH, 
where $L_{\rm Edd}$ is the Eddington luminosity. 
The spectral and timing properties are similar to those of BHBs in the high/soft state, and 
the spectrum in 1991 (the lowest luminosity case) seems to be that in the critical luminosity level for high to low state. 
Adopting the critical luminosity of $L/L_{\rm Edd}$=0.01--0.04 (e.g., \cite{Maccarone2003}), 
it supports the idea that X1755$-$338 is a low mass BH at a long distance. 

If X1755$-$338 is a BHB with $M_{\rm BH}\ge3 M_{\odot}$, the inclination angle of $<$70$^{\circ}$, 
and the distance is 4--9 kpc, 
another possible scenario is that X1755$-$338 is a rotating BH (the Kerr BH) 
because the last stable orbit around the rotating BH is thought to be smaller than the Schwartzshild radius (e.g., \cite{McClintock2006}). 
In fact, we fitted the spectra with a model consisting of an emission from the accretion disk 
for the Kerr BH ({\tt kerrbb} in XSPEC, \cite{Li2005}) 
and a thermal Comptonized continuum ({\tt nthcomp})
assuming $M_{\rm BH}=$ 3$M_{\odot}$, $\kappa=$ 1.7, 
a temperature of seed photon to be Compton-scattered of 1 keV, 
and an electron temperature of the {\tt nthcomp} model of 100 keV. 
Consequently, we found that the model can represent the spectra with 
the specific angular momentum, $a$ ($-1\le a <1$), of 0.55--0.84 
for the inclination angle of 70$^{\circ}$--60$^{\circ}$ 
at the distance of 9 kpc, respectively, and $a$=0.93--0.99 for 70$^{\circ}$--60$^{\circ}$ at 4 kpc, respectively. 
  

\begin{ack}
We thank Prof. Ken Ebisawa for helpful comments.
This research has made use of the DARTS system provided by ISAS/JAXA. 
\end{ack}


\end{document}